\documentstyle[12pt]{article}

\begin{document}
\begin{titlepage}

\hfill{UM-P-94/68}

\hfill{RCHEP-94/20}

\hfill{IP-ASTP-12}

\vskip 1 cm

\centerline{\bf \large New physics motivated by the low energy approach}
\centerline{\bf \large to electric charge quantization}

\vskip 1.5 cm

\centerline{{\large H. Lew}\footnote{lew@phys.sinica.edu.tw}$^{(a)}$
{\large and R. R. Volkas}\footnote{U6409503@hermes.ucs.unimelb.edu.au}
\footnote{Address from July-December 1994: Department of Applied
Mathematics and Theoretical Physics, University of Cambridge,
Silver Street, Cambridge, England CB3 9EW}$^{(b)}$}

\vskip 1.0 cm
\noindent
\centerline{{\it (a) Institute of Physics, Academia Sinica,}}
\centerline{{\it Nankang, Taipei, Taiwan 11529, R.O.C.}}

\vskip 0.25cm
\noindent
\centerline{{\it (b) Research Centre for High Energy Physics,
School of Physics,}}
\centerline{{\it University of Melbourne, Parkville 3052, Australia.}}

\vskip 1.0cm

\centerline{Abstract}

\noindent

The low-energy approach to electric charge quantization predicts
physics beyond the minimal standard model. A model-independent
approach via effective Lagrangians is used examine the possible
new physics, which may manifest itself indirectly through
family-lepton--number violating rare decays.

\end{titlepage}

The observed quantization of electric charge has long been a profound puzzle in
physics. The fact that hydrogen atoms and neutrons are electrically neutral
(to within experimental precision) helps shape the physics of the everyday
world. Electric charge quantization is thus a most important fact of nature,
and our understanding would be seriously incomplete if we could not
fathom why all particles carry integer multiples of the down-quark charge.

Until the last few years, grand unification of the strong, weak and
electromagnetic interactions seemed to be the most likely way electric charge
quantization would eventually be understood. Unfortunately, grand unified
theories are difficult to test experimentally because of the extremely high
energy scales involved (typically $10^{14-16}$ GeV). It is possible some hint
of grand unification such as proton decay may surface at any time if we are
fortunate. However, even if proton decay were to be discovered it would really
only tell us that baryon number is not conserved, and there are many ways to
violate baryon number conservation without invoking grand unification. In
general, any such evidence we may find will be at best indirect and suggestive
rather than compelling.

It is therefore important to explore ways of understanding electric
charge quantization that do not involve physics at largely inaccessible energy
scales. In recent years, a simple approach to the problem based on the
classical and quantal gauge invariance of the Standard Model (SM) Lagrangian
has been explicated in the literature. This work has shown that the oft-quoted
proclamation that the SM sheds no light on electric charge quantization is
wrong, and it provides hope that this puzzle can be solved in the foreseeable
future. Importantly, the low-energy approach to electric charge quantization
predicts that physics beyond the minimal SM is required for our understanding
to be complete. The task of this paper is to introduce a model-independent
strategy via effective Lagrangians for thinking about what this new physcis
might be.

We will not repeat the precise details of the low-energy electric
charge quantization calculations here, because they can be easily accessed
through review articles and the original papers \cite{ecq}.
However, by way of reminder let us go through the main steps in the
analysis for the SM:
\vskip 2 mm

\noindent
(i) We first write
down the multiplet assignments of all the particles in the theory under the
non-Abelian SU(3)$_c \otimes$SU(2)$_L$ part of the gauge group. All of the weak
hypercharge or $Y$ quantum numbers are left as arbitrary parameters.

\vskip 2 mm

\noindent
(ii) We use the arbitrary normalization of U(1) charges
to rescale the hypercharge of
the Higgs doublet $\phi$ to be 1. This is purely a matter of convenience. We
then use SU(2)$_L$ gauge invariance to write the vacuum expectation value (VEV)
of $\phi$ in the conventional form $\langle\phi\rangle = (0,v)^T$. The electric
charge generator $Q$ is then defined to be that linear combination of $I_{3L}$
and $Y$ which annihilates $\langle\phi\rangle$, where $I_{3L}$ is the diagonal
generator of SU(2)$_L$. We find that $Q = I_{3L} + Y/2$ where we have again
assigned the arbitrary normalization of $Q$ to conform with convention. We now
know what electric charge is, so we can begin to discuss its quantization. This
means we have to establish the quantization of $Y$.

\vskip 2 mm

\noindent
(iii) Particle physics up to about 100 GeV is well described by the SM
Lagrangian. So we now write down this Lagrangian, which forces us to relate
some of the hypercharges of the fermions in order to ensure
the gauge invariance of the Yukawa interaction terms. In the case of the SM,
some arbitrary hypercharges remain.

\vskip 2 mm

\noindent
(iv) Gauge anomaly cancellation is now imposed in order to protect the gauge
symmetry against quantal breaking. This further constrains the
hypercharge parameters.

\vskip 2 mm

No further sensible constraints exist. If the above procedure were
enough to force all hypercharge parameters to take on unique values, we
would conclude that the construction of the theory would only be possible if
hypercharge and hence electric charge quantization were to hold. This is the
sense in which electric charge quantization would be understood. If all the
parameters were to turn out to be fixed, then a major feature of the theory
such
as gauge invariant Yukawa coupling terms or anomaly cancellation would have to
be
sacrificed if one wanted to dequantize electric charge. Since this would be too
high a price to pay, we would deem electric charge quantization to be
understood
by consistency with the rest of known particle physics. However, if the
procedure
above were to fail in determining all of the hypercharge parameters, then
electric charge quantization would not be a necessary consequence of the
construction of the model.

What happens in the construction of the minimal SM (that is, the SM without
right-handed neutrinos)? If only one generation of fermions is considered, then
the above procedure is sufficient to fix all of the hypercharge parameters.
Electric charge quantization is thus understood.
However, in the realistic case
of three generations, one hypercharge parameter remains undetermined. Electric
charge quantization is thus not understood, although the form of electric
charge
dequantization is severely constrained.

To be specific, weak hypercharge and hence electric charge can be dequantized
in three similar but mutually exclusive ways in the three-generation minimal
SM. This is conveniently expressed by writing the actual weak hypercharge of
the theory $Y$ as a linear combination of standard weak hypercharge $Y_{\rm
st}$
and another generator. The three allowed forms for $Y$ are
\begin{equation}
Y = Y_{\rm st} + \epsilon (L_e - L_{\mu})\quad {\rm or}\quad
Y = Y_{\rm st} + \epsilon (L_e - L_{\tau})\quad {\rm or}\quad
Y = Y_{\rm st} + \epsilon (L_{\mu} - L_{\tau}),
\end{equation}
where $\epsilon$ is the arbitrary parameter, and $L_{e,\mu,\tau}$ are the
family-lepton--number generators.

The presence of family-lepton--number differences can be easily understood {\it
a posteriori}. We know that the three-generation minimal SM has five U(1)
invariances which commute with SU(3)$_c \otimes$SU(2)$_L$. The generators of
these U(1) groups are: standard hypercharge $Y_{\rm st}$, baryon number $B$ and
the family-lepton--numbers $L_{e,\mu,\tau}$. Any anomaly-free linear
combination of these five charges can be chosen as the generator of the gauged
U(1) in the SM gauge group. Apart from $Y_{\rm st}$, the family-lepton--number
differences are the only anomaly-free combinations.\footnote{If cancellation
of the mixed hypercharge-gravitational anomaly is not imposed, then there are
other gauge anomaly-free combinations \cite{ecq}.}
Therefore nothing prevents
us from gauging any linear combination of $Y_{\rm st}$ and one of these
differences, and so a 1-parameter dequantization of actual weak hypercharge
results. Note that the three family-lepton--number differences are not mutually
anomaly-free, which is why there are three distinct 1-parameter solutions for
$Y$.

This analysis provides strong motivation for extending the minimal SM in such a
way that electric charge quantization can be understood through the steps
outlined above. The perspective cast by the preceeding paragraph provides the
simplest way of stating what characteristics this new physics should have. We
must end up with U(1)$_{Y_{\rm st}}$ being the only anomaly-free U(1)
invariance
of the Lagrangian. Our task is therefore to explicitly break $L_e-L_{\mu}$,
$L_e-L_{\tau}$ and $L_{\mu}-L_{\tau}$ without introducing any other
anomaly-free
Abelian invariances of the Lagrangian (such as $B-L$).

Several concrete suggestions for what this new physics might
be have been canvassed in the literature. For instance, the introduction of
three generations of Majorana right-handed neutrinos is sufficient \cite{babu}.
(Kobayashi-Maskawa--like mixing in the lepton sector in general explicitly
breaks
all of the family-lepton--number symmetries, but the right-handed neutrinos
must
also have a Majorana character to explicitly break $B-L$. Although $B-L$ is not
anomaly-free in the minimal SM, it is anomaly-free when three right-handed
neutrinos are added.) Alternatively, one can introduce only one right-handed
neutrino state which need not be Majorana \cite{foot}.
A different suggestion is that two
Higgs doublets be used to explicitly
break the troublesome $L_i-L_j$ symmetries \cite{sladkowski}.

There are many other candidates for the required new physics. Whatever the new
physics might be, we know that it must explicitly break all of the $L_i-L_j$
invariances. It therefore makes sense to perform a {\it model-independent}
analysis using effective Lagrangian techniques of all possible
higher-dimensional operators that explicitly break family-lepton--number
differences. The task of this paper is to begin such an analysis. We will
assume that low-energy physics can be described by an effective Lagrangian
written in terms of the fields of the minimal SM only (in particular, we will
exclude right-handed neutrinos from the low-energy world). Non-renormalizable
operators breaking $L_i-L_j$ will be constructed from these fields, and
experimentally relevant processes induced by these operators will be
identified and bounds given. We will draw conclusions about what the underlying
renormalizable extension of the minimal SM should look like whenever
appropriate.

The building blocks of our analysis are the fields of the minimal SM. Each
generation of fermions has the
$G_{\rm SM} = $SU(3)$_c\otimes$SU(2)$_L\otimes$U(1)$_{Y_{\rm st}}$ structure
\begin{eqnarray}
&\ell_L \sim (1,2)(-1),\quad e_R \sim (1,1)(-2),&\ \nonumber\\
&q_L \sim (3,2)(1/3),\quad u_R \sim (3,1)(4/3),\quad d_R \sim (1,3)(-2/3).&\
\end{eqnarray}
The gauge bosons have their usual transformation properties, while the Higgs
doublet $\phi$ is characterised by $\phi \sim (1,2)(1)$.

We will assume that the new physics can be assigned an energy scale $\Lambda$
which is higher than the electroweak scale of 300 GeV. The scale $\Lambda$
will provide the ultraviolet cut-off for the effective theory, and the
influence
of all non-renormalizable operators on low-energy physics will be
suppressed by powers of some typical SM energy or mass divided by $\Lambda$.
Dimension-5 operators will have a $1/\Lambda$ suppression in the Lagrangian,
while dimension-6 operators will be suppressed by $(1/\Lambda)^2$, and so on.

We will restrict our analysis to dimension-5 and -6 operators in this paper.
This is logical because of the general expectation that the higher the
dimension of the operator the more it is suppressed by powers of
mass/$\Lambda$. However, we should be aware of important ways this procedure
may be misleading. First, it is possible that a symmetry of the underlying
renormalizable theory may forbid all operators from dimension-5 up to some
higher dimension, say dimension-7 by way of illustration. In that case, the new
physics responsible for ensuring charge quantization  will manifest itself
first at the dimension-8 level, and the analysis of this paper will be
irrelevant. Although this is an interesting possibility, we will for
simplicity not focus on it here. Second, although the underlying dynamics may
generate dimension-5 and -6 terms, they may necessarily come in with a larger
suppression factor than the naive ``mass over $\Lambda$ to some power''
expectation. For instance, the topology of Feynman diagrams can prevent certain
dimension-6 operators being generated at tree-level by {\it any} underlying
gauge theory \cite{aew}.
In this case there is always an additional loop suppression factor
of at least $1/16\pi^2$ to the amplitude of the process, provided the
underlying physics does not have a non-perturbative way of generating
the dimension-6 operator in question\footnote{This amounts to
a suppression factor of $1/4\pi$ for $\Lambda$.}. It is then possible
for dimension-7 and -8 operators to be more important than some
dimension-6 operators.

The tedious task of writing down all dimension-5 and -6 operators for the
minimal SM has been performed \cite{lag}.
We will be concerned only with the subset that violates $L_i-L_j$.

Only one set of dimension-5 operators can be constructed under the stated
assumptions, and since they happen to break family- (and total-)
lepton--number they are relevant. They are given by
\begin{equation}
O^5_{ij} = \overline{(\ell_{iL})^c} i\tau_2 \phi\ \ell_{jL} i\tau_2 \phi
\end{equation}
together with their hermitian conjugates, where $i,j=1,2,3$ are generation
indices. Each of these operators breaks two of the $L_i-L_j$ charges and
preserves another. Therefore at least two suitably chosen operators from the
$O^5_{ij}$ set must be simultaneously present. After electroweak symmetry
breakdown, these operators induce Majorana terms for the left-handed neutrinos.
The coefficients of these operators can be very severely bounded by
experiment. The most stringent bound applies to the coefficient
$a_{11}/\Lambda$ of $O^5_{11}$ because this operator induces a Majorana mass
for the left-handed electron-neutrino, given by
\begin{equation}
m^{\rm Majorana}_{\nu_e} = a_{11} {v^2 \over \Lambda}.
\end{equation}
The experimental upper bound is about 1 eV, which leads to the constraint
\begin{equation}
\Lambda > a_{11} \times 10^{14}\ {\rm GeV}.
\label{dim5bound}
\end{equation}
Adopting the general expectation that $a_{11} \simeq 1$, we see that $\Lambda$
should be greater than the very large value of about $10^{14}$ GeV. The other
operators in this set will not provide such stringent bounds, but the
typical lower bounds on $\Lambda$ will be very high nevertheless.

This is an unsatisfactory result with regard to charge quantization, because we
were after all endeavouring to find the required new physics at relatively
low-energies like 1 TeV. So, the underlying dynamics must either forbid these
dimension-5 terms, or suppress them sufficiently. The obvious way the
underlying renormalizable theory could forbid these terms is to insist that
total-lepton--number $L$ (or some linear combination of baryon and lepton
number
such as $3B+L$) be conserved. This is interesting information.

Could the coefficients somehow receive a large enough suppression? We can give
a
qualified answer of ``yes'' to this question. It is probably fair to say that
the most natural candidate thus far proposed as new physics to ensure
low-energy charge quantization is the addition of three Majorana right-handed
neutrinos. In particular, it is then natural to use the see-saw mechanism to
explain why left-handed neutrinos are so light. Recall that all of the mass
eigenstate neutrinos in the see-saw model are Majorana, including the light
left-handed neutrinos which have masses generically given by $m_D^2/M$, where
$m_D$ is a Dirac mass and $M$ is a Majorana mass with $M \gg m_D$. Furthermore,
the see-saw mass matrix induces precisely the dimension-5 operators we have
been
discussing once the heavy right-handed Majorana neutrinos are integrated out!
Why is this such a popular candidate given the pessimistic result of
Eq.(\ref{dim5bound})? The answer is that we generally expect the Dirac masses
$m_D$ to be many orders of magnitude smaller than the electroweak scale $v$,
simply because this is so for all observed quark and charged-lepton masses
except
for the top quark. The Yukawa coupling constants for quarks and charged-leptons
are unexplained small numbers in the SM, and we simply assume that the Yukawa
coupling constants involved in neutrino Dirac masses are similarly unexplained
small parameters. In effective Lagrangian language, this means that the
dimensionless coefficients $a_{ij}$ are actually many orders of magnitude
smaller than 1. This means the lower bound on $\Lambda$ can be lowered to a
respectable level. For instance, if we desire $\Lambda \simeq 1$ TeV then we
need $a_{11} \simeq 10^{-11}$. Since in the see-saw model $a_{11}$ is the
product of two Dirac neutrino Yukawa coupling constants, we see that values
comparable in smallness to the electron Yukawa coupling constant are needed.
This ``explanation'' of the suppression of the $a_{ij}$'s is of course highly
unsatisfactory, but this just reflects the highly unsatisfactory status of
fermion mass generation in the SM.

If Majorana right-handed neutrinos constitute the new physics then our story
ends with the dimension-5 terms, and the smallness of the neutrino Dirac Yukawa
coupling constants is left to be explained by a hypothetical theory of flavour.
In this paper, we will instead consider other possible explanations. To forbid
the $O^5_{ij}$ operators we will suppose that total-lepton--number is conserved
by the underlying theory\footnote{Some linear combination of $B$ and $L$
which is conserved will also forbid these operators.} and we now move on
to dimension-6 terms.

There are many dimension-6 operators which conserve $L$ and $B$ but violate
$L_i-L_j$. They can be gleaned from the list given in Buchm\"uller and Wyler
in Ref. \cite{lag}. Using their notation they are:

\vskip 2 mm

\noindent
{\it Four-fermion Operators}
\begin{eqnarray}
&O^{(1)}_{\ell\ell} = {1 \over 2}(\overline{\ell}_L \gamma_{\mu}
\ell_L)(\overline{\ell}_L \gamma^{\mu} \ell_L),\qquad
O^{(3)}_{\ell\ell} = {1 \over 2}(\overline{\ell}_L \gamma_{\mu}\tau^I
\ell_L)(\overline{\ell}_L \gamma^{\mu}\tau^I \ell_L),&\ \nonumber\\
&O^{(1)}_{\ell q} = {1 \over 2}(\overline{\ell}_L \gamma_{\mu}
\ell_L)(\overline{q}_L \gamma^{\mu} q_L),\qquad
O^{(3)}_{\ell q} = {1 \over 2}(\overline{\ell}_L \gamma_{\mu}\tau^I
\ell_L)(\overline{q}_L \gamma^{\mu}\tau^I q_L),&\ \nonumber\\
&O_{ee} = {1 \over 2}(\overline{e}_R \gamma_{\mu} e_R)(\overline{e}_R
\gamma^{\mu} e_R),\qquad
O_{eu} = (\overline{e}_R \gamma_{\mu} e_R)(\overline{u}_R
\gamma^{\mu} u_R),&\ \nonumber\\
&O_{ed} = (\overline{e}_R \gamma_{\mu} e_R)(\overline{d}_R
\gamma^{\mu} d_R),\qquad
O_{\ell e} = (\overline{\ell}_L e_R)(\overline{e}_R \ell_L),&\ \nonumber\\
&O_{\ell u} = (\overline{\ell}_L u_R)(\overline{u}_R \ell_L),\qquad
O_{\ell d} = (\overline{\ell}_L d_R)(\overline{d}_R \ell_L),&\ \nonumber\\
&O_{qe} = (\overline{q}_L e_R)(\overline{e}_R q_L),\qquad
O_{qde} = (\overline{\ell}_L e_R)(\overline{d}_R q_L),&\ \nonumber\\
&O_{\ell q} = (\overline{\ell}_L e_R)(\overline{q}_L e_R);&\
\end{eqnarray}

\vskip 2 mm

\noindent
{\it Operators with Fermions and Vector Bosons}
\begin{eqnarray}
&O_{\ell W} = i\overline{\ell}_L \tau^I \gamma_{\mu} D_{\nu} \ell_L W^{I\mu\nu}
\qquad O_{\ell B} = i \overline{\ell}_L \gamma_{\mu} D_{\nu} \ell_L
B^{\mu\nu},&\ \nonumber\\
&O_{eB} = i \overline{e}_R \gamma_{\mu} D_{\nu} e_R B^{\mu\nu};&\
\end{eqnarray}

\vskip 2 mm

\noindent
{\it Operator with Fermions and Scalars}
\begin{equation}
O_{e\phi} = (\phi^{\dagger} \phi)(\overline{\ell}_L e_R \phi);
\end{equation}

\vskip 2 mm

\noindent
{\it Operators with Fermions, Scalars and Vector Bosons}
\begin{eqnarray}
&O^{(1)}_{\phi\ell} = i(\phi^{\dagger} D_{\mu} \phi)(\overline{\ell}_L
\gamma^{\mu} \ell_L),\qquad
O^{(3)}_{\phi\ell} = i(\phi^{\dagger} D_{\mu} \tau^I \phi)(\overline{\ell}_L
\gamma^{\mu} \tau^I \ell_L),&\ \nonumber\\
&O_{\phi e} = i(\phi^{\dagger} D_{\mu} \phi)(\overline{e}_R \gamma^{\mu}
e_R),\qquad
O_{D_e} = (\overline{\ell}_L D_{\mu} e_R)D^{\mu} \phi,&\ \nonumber\\
&O_{\overline{D}_e} = (D_{\mu}\overline{\ell}_L e_R)D^{\mu} \phi,\qquad
O_{eW} = (\overline{\ell}_L \sigma^{\mu\nu} \tau^I e_R)\phi W^I_{\mu\nu},&\
\nonumber\\
&O_{\ell eB} = (\overline{\ell}_L \sigma^{\mu\nu} e_R) \phi B_{\mu\nu}.&\
\end{eqnarray}
Generation indices have been suppressed in these equations, while $W^{I\mu\nu}$
and $B^{\mu\nu}$ are the field strength tensors for SU(2)$_L$ and U(1)$_{Y{\rm
st}}$ respectively.\footnote{Note that the symbol ``$O_{eB}$'' appears in the
Buchm\"uller and Wyler list in both Eq.(3.31) and Eq.(3.60). We have renamed
the
last of these as ``$O_{\ell eB}$''.}

We must now look at which $L_i-L_j$ violating processes these operators can
induce. The Particle Data Group's Review of Particle Properties \cite{pdg}
lists bounds on the several dozen family-lepton--number processes that have
been looked for experimentally.
It is interesting to classify these processes by examining how
many units of $L_e-L_{\mu}$, $L_e-L_{\tau}$ and $L_{\mu}-L_{\tau}$ they
violate. We will denote the number of units violated by $\Delta_{e\mu}$,
$\Delta_{e\tau}$ and $\Delta_{\mu\tau}$, respectively. Let us make a few
preliminary observations: (i) Most of these processes have nonzero values for
each of the $\Delta_{ij}$'s. For instance, the process $\mu \to e\gamma$ has
$\Delta_{e\mu} = 2$, $\Delta_{e\tau} = 1$ and $\Delta_{\mu\tau} = 1$.
Therefore, if this rare decay is ever observed to happen we will be able to
conclude on the basis of this single process that all family-lepton--number
differences are not conserved and that new physics associated with the charge
quantization problem has been found.\footnote{However,
we could not be sure that some other U(1) generator such as $B-L$ was not
rendered anomaly-free according to the as yet unknown underlying theory.
To be sure of this we would need much more experimental information so that we
could construct the entire theory.}
(ii) Like $\mu \to e\gamma$, the majority of
the processes listed have one of the $\Delta_{ij}$'s equal to 2, with the other
two equal to 1. (iii) Two rare decays of the tau lepton, $\tau^- \to
e^+\mu^-\mu^-$ and $\tau^- \to
\mu^+e^-e^-$, conserve one of the family-lepton--number differences
($L_e-L_{\tau}$ and $L_{\mu}-L_{\tau}$ respectively). Therefore the observation
of one of these decays in isolation would not necessarily signal the presence
of new physics enforcing charge quantization, even though there would certainly
be new physics.

Let us now list a representative selection of interesting processes according
to
their pattern of $L_i-L_j$ violation:

\vskip 5 mm

\noindent
{\bf A. $\Delta_{e\mu} = 2$, $\Delta_{e\tau} = 1$, $\Delta_{\mu\tau} = 1$}
\begin{eqnarray}
&B(Z \to e^{\pm}\mu^{\mp}) < 2.4 \times 10^{-5},\quad
B(\mu^- \to e^- \gamma) < 5 \times 10^{-11},&\ \nonumber\\
&B(\mu^- \to e^- e^+ e^-) < 1.0 \times 10^{-12},\quad
B(\pi^0 \to \mu^+ e^-) < 1.6 \times 10^{-8},&\ \nonumber\\
&B(K^+ \to \pi^+ e^- \mu^+) < 2.1 \times 10^{-10},\quad
B(K^0_L \to e^{\pm} \mu^{\mp}) < 9.4 \times 10^{-11}.&\
\end{eqnarray}

\vskip 5 mm

\noindent
{\bf B. $\Delta_{e\mu} = 1$, $\Delta_{e\tau} = 2$, $\Delta_{\mu\tau} = 1$}
\begin{eqnarray}
&B(Z \to e^{\pm} \tau^{\mp}) < 3.4 \times 10^{-5},\quad
B(\tau^- \to e^- \gamma) < 2.0 \times 10^{-4},&\ \nonumber\\
&B(\tau^- \to e^- \pi^0) < 1.4 \times 10^{-4},\quad
B(\tau^- \to e^- \rho) < 3.9 \times 10^{-5},&\ \nonumber\\
&B(\tau^- \to e^- e^+ e^-) < 2.7 \times 10^{-5},\quad
B(\tau^- \to e^- \mu^+ \mu^-) < 2.7 \times 10^{-5}.&\
\end{eqnarray}

\vskip 5 mm

\noindent
{C. $\Delta_{e\mu} = 1$, $\Delta_{e\tau} = 1$, $\Delta_{\mu\tau} = 2$}
\begin{eqnarray}
&B(Z \to \mu^{\pm} \tau^{\mp}) < 4.8 \times 10^{-5},\quad
B(\tau^- \to \mu^- \gamma) < 5.5 \times 10^{-4},&\ \nonumber\\
&B(\tau^- \to \mu^- \pi^0) < 8.2 \times 10^{-4},\quad
B(\tau^- \to \mu^- \rho) < 3.8 \times 10^{-5},&\ \nonumber\\
&B(\tau^- \to \mu^- \mu^+ \mu^-) < 1.7 \times 10^{-5},\quad
B(\tau^- \to \mu^- e^+ e^-) < 2.7 \times 10^{-5}.&\
\end{eqnarray}

\vskip 5 mm

There are some processes which do not fall into any of these categories. We
have already discussed $\tau^- \to e^+\mu^-\mu^-$ and $\tau^- \to \mu^+e^-e^-$.
There is also the result $B(\mu^- \to e^- \nu_e \overline{\nu}_{\mu}) < 1.8
\times 10^{-2}$ which obeys $\Delta_{e\mu} = 4$, $\Delta_{e\tau} =
\Delta_{\mu\tau} = 2$.

The processes in category A provide the most stringent bounds on $\Lambda$. The
most severe constraint comes from $\mu \to e\gamma$. This decay can be induced
by the operators $O_{\ell W}$, $O_{\ell B}$, $O_{eB}$, $O_{eW}$ and
$O_{\ell eB}$, yielding typically that
\begin{equation}
\Lambda> 10^7\ {\rm GeV}.
\end{equation}
The next most severe constraint comes from $\mu \to 3e$ which can be induced by
$O^{(1)}_{\ell\ell}$, $O^{(3)}_{\ell\ell}$ and $O_{ee}$. The typical bound is
\begin{equation}
\Lambda > 10^5\ {\rm GeV}.
\end{equation}
The decays $K^+ \to \pi^+ e^- \mu^+$ and $K_L^0 \to \mu^{\pm} e^{\mp}$ both
yield $\Lambda > 5 \times 10^4$ GeV or so. The LEP bound on $Z \to
e^{\pm}\mu^{\mp}$ implies that $\Lambda > 1$ TeV, while $\pi^0 \to
\mu^{\pm}e^{\mp}$ implies the very weak bound that $\Lambda > 90$ GeV.
(The various effective operators containing both quarks and leptons contribute
to the processes above that involve hadrons.)

The operator $O_{e\phi}$ induces flavour-changing vertices between the physical
Higgs boson and the charged leptons. At tree-level this will contribute to
$\mu \to 3e$, while at one-loop level it will contribute to $\mu \to e\gamma$.
However, we find the bounds on $\Lambda$ due to these Higgs boson effects
to be weaker than those derived above.

Clearly, if category A processes are responsible for enforcing charge
quantization, then the scale of the new physics is typically at the rather high
value of $10^7$ GeV. This sort of new physics will therefore be difficult to
explore directly. Of course, the fact that the bound on $\Lambda$ from decays
like $\mu \to e \gamma$ is so severe reflects our ability to
do very high statistics searches for this decay mode. Therefore we may well
observe a nonzero rate for this process as statistics improve further, despite
a
high value for $\Lambda$. This would serve as a dramatic manifestation of the
sought after new physics. However, rare decay searches are indirect rather than
direct explorations of physics beyond the SM. Ideally, we would like to be able
to experiment on the totality of the new physics and not just on its subtle
low-energy effects. This would require studying collisions at $\Lambda$
energies. We therefore conclude that if the non-standard physics induces
category A processes at the dimension-6 level then we can realistically only
ever expect to study its indirect effects.

According to the analysis of Ref. \cite{aew}, the dimension-6 operators
inducing $f \to f'\gamma$ cannot be generated at tree-level by any underlying
gauge theory.
Therefore the bound $\Lambda > 10^7$ GeV should in this case be reduced by
about $1/4\pi$, since the coefficient of the operator will necessarily have
a loop suppression factor. This brings the $\mu \to e\gamma$ lower bound on
$\Lambda$ into the 800 TeV regime, which is still very high
thus requiring a post-LHC machine
to study the new physics directly. Furthermore, if the new physics has
non-perturbative or perhaps even non-gauge character then this argument becomes
moot.

Let us now turn to category B. The most severe constraint comes once again from
radiative lepton decay. The bound on $\tau \to e\gamma$ yields
\begin{equation}
\Lambda > 40\ {\rm TeV}.
\end{equation}
Other relevant processes are $\tau \to 3e$, $\tau \to e2\mu$, $Z \to e\tau$ and
$\tau \to \rho e$ which all imply that $\Lambda >$ 1-2 TeV or so, while the
lower bound from $\tau \to \pi^0 e$ is a little lower than a TeV. Although 40
TeV is still a little high, the
$1/4\pi$ suppression that occurs if the new physics is perturbative yields
$\Lambda > 3$ TeV from
$\tau \to e\gamma$. Category C bounds are roughly
the same as those from category B. So, it is quite possible for category B
and C physics to exist at TeV scale energies, which is a pleasing conclusion.

What have we learned from these observations? First, the underlying dynamics is
likely to respect total-lepton--number (or some linear combination of $B$
and $L$) conservation so that no dimension-5
terms are induced. However, this is far from being a rigorous requirement, as
the example of the see-saw model demonstrates. Second, if the new
dynamics is to operate at LHC energies, then the $\Delta_{e\mu}=2$,
$\Delta_{e\tau}=1$,
$\Delta_{\mu\tau}=1$ class of processes must be prevented from occuring at the
dimension-6 level. This can happen if the underlying dynamics conserves
some linear combination of $L_\mu$ (or $L_e$) and $L_\tau$
while at the same time breaking
$\Delta_{e\mu}$, $\Delta_{e\tau}$ and $\Delta_{\mu\tau}$.
In any case, the observation of such a process would nonetheless be an
exciting indirect manifestation of non-standard dynamics. Third, if processes
respecting $\Delta_{e\mu}=1$, $\Delta_{e\tau}=2$,
$\Delta_{\mu\tau}=1$ or $\Delta_{e\mu}=1$, $\Delta_{e\tau}=1$,
$\Delta_{\mu\tau}=2$ are discovered in the near future, they may be a signal
of new dynamics at the TeV scale. Fourth, it is interesting to also contemplate
underlying models which do not generate any $L_i-L_j$ violating dimension-5 and
-6 terms. This will serve to lower the bound on the scale of new physics
further, and if the model is constructed correctly will allow category A
processes like $\mu \to e\gamma$ to be induced by TeV scale dynamics. At any
rate, effective operators provide a systematic and useful way to classify the
phenomenological consequences of underlying theories, and they can even provide
hints as to how to build these models.

\vskip 1.5cm
\leftline{\bf Acknowledgements}
We would like to thank R. Foot for some discussions.
RRV would like to thank Professor H.-Y. Cheng and the Institute of Physics at
Academia Sinica in Taiwan, and also Professor J. C. Taylor and DAMTP at
Cambridge University for their kind hospitality while portions of this
work were done.

\end{document}